\documentclass[floats,10pt,onecolumn]{revtex4}
\usepackage{enumerate}
\usepackage{xcolor}
\definecolor{DarkGray}{rgb}{0.1,0.1,0.5}
\usepackage[colorlinks=true,breaklinks, linkcolor= DarkGray,citecolor= DarkGray,urlcolor= DarkGray]{hyperref}
\usepackage{url}
\usepackage{verbatim}
\usepackage{amsmath}
\usepackage{latexsym}
\usepackage{revsymb}
\usepackage{epsfig}
\usepackage{pstricks}
\usepackage{ifthen}
\usepackage{mathrsfs}

\usepackage{natbib}
\usepackage{amsfonts}
\usepackage{amsmath}
\usepackage{amssymb}
\usepackage{subfigure}
\usepackage{amsthm}
\usepackage{graphicx}
\newcommand{\appref}[1]{\hyperref[#1]{{Appendix~\ref*{#1}}}}
\newcommand{\be}{\begin{eqnarray} \begin{aligned}}
\newcommand{\ee}{\end{aligned} \end{eqnarray} }
\newcommand{\benn}{\begin{eqnarray*} \begin{aligned}}
\newcommand{\eenn}{\end{aligned} \end{eqnarray*} }

\newcommand{\bc}{\begin{center}}
\newcommand{\ec}{\end{center}}

\newcommand{\myacknowledgments}{\begin{center}{\bf Acknowledgments}\end{center}\par}

\usepackage{amsfonts}

\def\01{\{0,1\}}

\newcounter{protoCount}
\newcounter{protoList}
\newsavebox{\tmpbox}
\newlength{\protobox}

\newcommand{\Bob}{\text{Bob}}

\usepackage{amsmath}
\usepackage{amssymb}

\begin{document}

\title{Information Causality is a Special Point in the Dual of the Gray-Wyner Region}
\author{Salman Beigi\email{salman.beigi@gmail.com}}
\affiliation{School of Mathematics,
 Institute for Research in Fundamental Sciences (IPM),
 Tehran, Iran}
\author{Amin Gohari \email{amin.aminzadeh@gmail.com}}
\affiliation{ Department of Electrical Engineering, Sharif University of Technology, Tehran, Iran\\
School of Computer Science,
 Institute for Research in Fundamental Sciences (IPM), Tehran, Iran
}

\begin{abstract}
Information Causality contributes to the program of deriving fundamentals of quantum theory from information theoretic principles.
It puts restrictions on the amount of information learned by a party (Bob) from the other party (Alice) in a one-way communication scenario: Bob receives an index $b$, and after a one-way communication from Alice, tries to recover the $b$-th bit of Alice's input.
In this manuscript we note that Information Causality is indeed a point in the dual of the Gray-Wyner region in the case where Alice's input bits are chosen independently at random. The main motivation of this work was to study this connection in the case of correlated input bits. Our approach to forge this connection was to propose a new postulate called Accessibility of Mutual Information (AMI) on the underling physical theory. This approach, explained in the first version of this paper, fails since AMI is not satisfied by the quantum theory.

\end{abstract}
\maketitle

Let us start with a brief description of the game of Information Causality (IC). Alice receives the bit-string $\overrightarrow{a}=(a_1,\dots, a_N)$ chosen according to some distribution $p(\overrightarrow{a})$, and Bob gets an index $1\leq b\leq N$ chosen uniformly at random. Bob's goal is to output $a_b$ upon receiving a classical message $x$ from Alice. Assuming that $\beta_i$ is Bob's guess of $a_i$ when $b=i$, and some notion of `mutual information' satisfying certain properties is available, IC states that if $p(\overrightarrow{a})$ is an i.i.d.\ distribution then~\cite{IC}
\begin{align}\label{eq:IC}
H(x) \geq \sum_{i=1}^N I(a_i ; \beta_i \vert b=i).
\end{align}
This inequality puts restrictions on feasible non-local correlations of the underlying physical theory.

When a consistent notion of entropy is also available one can rewrite this inequality in terms of entropies as follows \footnote{In~\cite{Short11} and~\cite{Renner11} this inequality has been derived from some postulates directly on the entropy rather than mutual information.
}:
\begin{align}\label{eq:IC-entropy-form-1}
H(x) + \sum_{i=1}^N H(a_i \vert \beta_i, b=i) \geq H(\overrightarrow{a}).
\end{align}
One can easily verify that this inequality still holds ever if we remove the assumption that $p(\overrightarrow{a})$ is an i.i.d.\ distribution. But letting $p(\overrightarrow{a})$ be arbitrary the question is whether we can prove other inequalities to restrict plausible correlations in the underlying physical theory.

Consider the game of IC in its completely classical form. Classicality allows us to assume that there are $N$ Bobs instead of one. We denote these $N$ Bobs by $\Bob_1, \dots, \Bob_N$. The goal of $\Bob_i$ is to find $a_i$. Moreover, we may assume that classical correlation, namely shared randomness, is indeed shared amongst Alice and all Bobs, and all of them receive the message $x$ from Alice. Then the first term $H(x)$ of \eqref{eq:IC-entropy-form-1} is the amount of information that is sent to all Bobs; the second term $H(a_i \vert \beta_i, b=i)$ expresses the remaining uncertainty of $\Bob_i$ about $a_i$. We can interpret this as the average number of extra bits that Alice needs to privately send to $\Bob_i$ to enable the recovery of $a_i$ by this party if they were to play multiple copies of this game in parallel (the Slepian-Wolf theorem). Since $\Bob_1, \dots, \Bob_N$ altogether can recover the string $\overrightarrow{a}$, the total flow of information from Alice should be at least $H(\overrightarrow{a})$ by the cut-set bound. That is, the sum of the terms on the left hand side of \eqref{eq:IC-entropy-form-1} should dominate $H(\overrightarrow{a})$. This gives a new proof of IC in the classical world.

\begin{figure}
\includegraphics[width=4.3in]{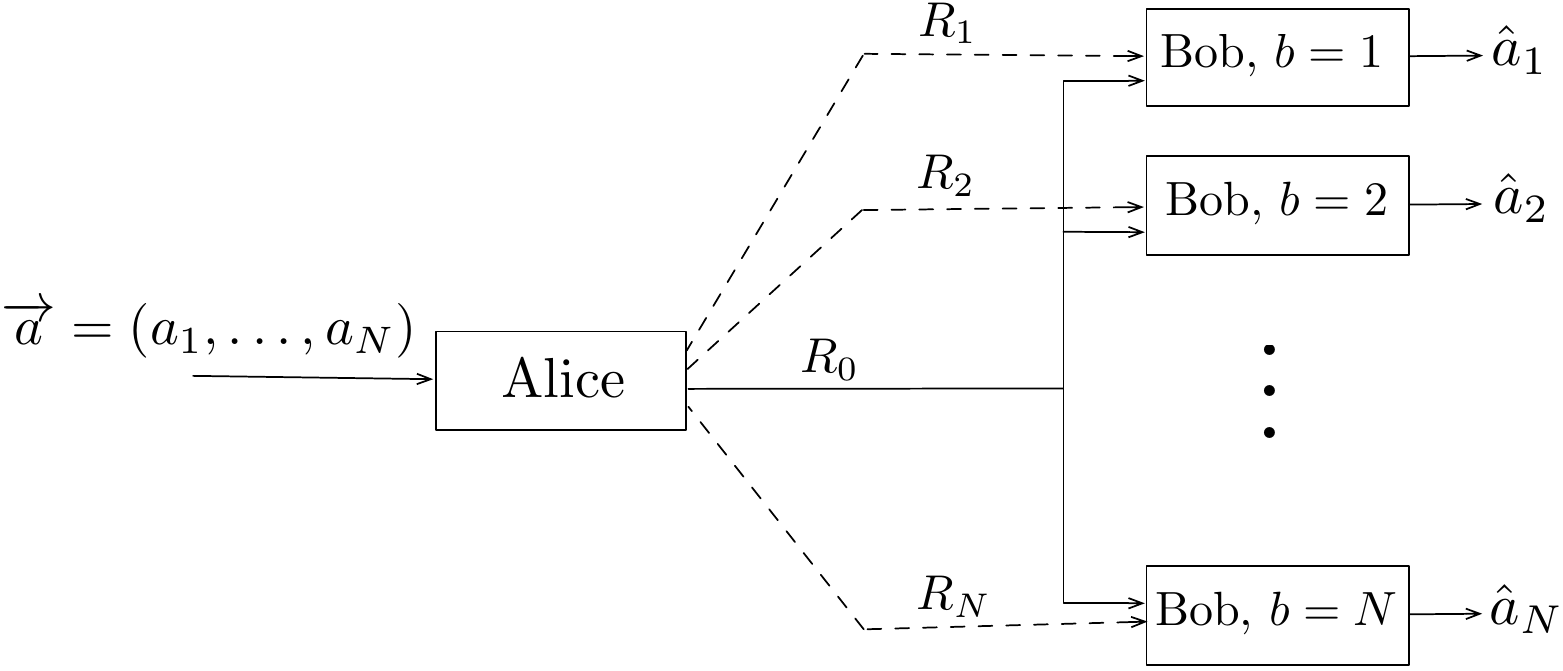}
\caption{The Gray-Wyner game consists of $N+1$ players, Alice and $N$ Bobs who are indexed by $b=1, \dots, N$. Alice receives the i.i.d.\ copies of $(a_1, \dots, a_N),$ chosen according to $p(a_1, \dots, a_N)$, sends public information at rate $R_0$ to all Bobs and private information at rate $R_i$ to $\Bob_i$. The goal of $\Bob_i$ is to recover $a_i$.
}\label{fig:gray-wyner}
\end{figure}

The above game among Alice and the multiple copies of Bob has a similar setup to the Gray-Wyner problem~\cite{GrayWyner} defined as follows. Alice sends a public message $x$ to all Bobs and afterwards a private message to each $\Bob_i$. The goal of $\Bob_i$ is to recover $a_i$ with a vanishing probability of error (see Fig.~\ref{fig:gray-wyner}). Let $R_0$ denote the information rate of the public message, and $R_1, \dots, R_N$ denote the rate of the private messages to $\Bob_1, \dots, \Bob_N$. The Gray-Wyner region explicitly characterizes the set of tuples $(R_0, R_1, \dots, R_N)$ for which it is possible to satisfy the demands of $\Bob_1, \dots, \Bob_N$. By the above discussion the rates $R_0=H(x)$ and $R_i=H(a_i\vert \beta_i, b=i)$ have to lie in the Gray-Wyner region when the IC game is played in the classical world.

In the special case where $p(\overrightarrow{a})=p_1(a_1)\cdots p_N(a_N)$ the Gray-Wyner region is completely characterized by inequalities 
\begin{align}\label{eq:gw-region}
R_0 + \sum_{i\in S}  R_i \geq H(\overrightarrow{a}_S),
\end{align}
for all subset $S\subseteq \{1, \dots, N\}$. By \eqref{eq:IC-entropy-form-1} these inequalities holds for $R_0=H(x)$ and $R_i=H(a_i\vert \beta_i, b=i)$ when the game of IC is played in an arbitrary physical theory satisfying some postulates. So IC in this special case can be stated in term of the Gray-Wyner region. Now the question is what if we remove the restriction on $p(\overrightarrow{a})$ and let it be arbitrary. In this case the Gray-Wyner region is determined by infinitely many inequalities and \eqref{eq:gw-region} describes only finitely many of them. The question is whether those inequalities still hold for $R_0=H(x)$ and $R_i=H(a_i\vert \beta_i, b=i)$ when the game of IC is played with an arbitrary $p(\overrightarrow{a})$ or not. 

In the first version of this paper available on arXiv (also presented at QIP2012) this question is answered in an affirmative way assuming a postulate on the underlying physical theory called Accessibility of Mutual Information (AMI). Although our arguments in this direction are correct, the quantum theory does not satisfy AMI, so our answer to this question does not seem fair. Here we report that all our approaches to solve this problem have been failed and this question is still open \footnote{This problem is indirectly attacked in \cite{BA} as well.}.

\myacknowledgments
The authors are thankful to Aram Harrow, Patrick Hayden and especially Andreas Winter for letting us know about our mistake during QIP2012.

\end{document}